\documentclass{article}
\usepackage{ijcai24}

\usepackage{times}
\usepackage{soul}
\usepackage{url}
\usepackage[hidelinks]{hyperref}
\usepackage[utf8]{inputenc}
\usepackage[small]{caption}
\usepackage{graphicx}
\usepackage{amsmath}
\usepackage{amsthm}
\usepackage{booktabs}
\usepackage{algorithm}
\usepackage{algorithmic}
\usepackage[switch]{lineno}

\usepackage{afterpage} 
\usepackage{booktabs} 
\usepackage{caption}
\usepackage{makecell} 
\usepackage{amssymb}
\usepackage{amsmath}

\usepackage{mdframed}
\usepackage{fancyvrb}
\usepackage[most]{tcolorbox}

\title{Geography-Aware Large Language Models for Next POI Recommendation}
\author{
    Zhao Liu\textsuperscript{\rm 1}, 
    Wei Liu\textsuperscript{\rm 1}, 
    Huaijie Zhu\textsuperscript{\rm 1}, 
    Jianxing Yu\textsuperscript{\rm 1}, 
    Jian Yin\textsuperscript{\rm 1}, \\
    Wang-Chien Lee\textsuperscript{\rm 2}, 
    Shun Wang\textsuperscript{\rm 1} \\
    \textsuperscript{\rm 1}Sun Yat-sen University \\
    \textsuperscript{\rm 2}The Pennsylvania State University \\
    \{liuzh368, liuw259, zhuhuaijie, yujx26, issjyin, wangsh363\}@mail.sysu.edu.cn, \\
    wlee@cse.psu.edu
}

\begin{document}

\maketitle

\begin{abstract}

The next Point-of-Interest (POI) recommendation task, aiming to predict users’ next destination based on their historical movement data, plays a crucial role in location-based services and personalized applications. Accurate next POI recommendation relies heavily on modeling geographic information and POI transition relations, which are critical for capturing spatial dependencies and user movement patterns. While Large Language Models (LLMs) demonstrate strong semantic understanding and contextual modeling, their applications to spatial tasks such as the next POI recommendation face significant challenges: (1) the infrequent occurrence of specific GPS coordinates hinders effective and precise spatial context modeling, and (2) the lack of POI transition knowledge limits LLMs' understanding of potential POI-POI relationships. We therefore propose Geography-Aware Large Language Model (GA-LLM), a novel framework that enhances LLMs with two specialized modules: i) Geographic Coordinate Injection Module (GCIM) transforms GPS coordinates into meaningful spatial representations using a hierarchical and Fourier positional encoding, enabling GA-LLM to effectively capture geographic features with multiple granularities and perspectives. ii) POI Alignment Module (PAM) integrates POI transition relations into the LLM’s semantic space, allowing it to infer global POI relationships and explore previously unseen POIs for users. Experiments on three datasets show the state-of-the-art performance of GA-LLM. The source code is available at: \href{https://anonymous.4open.science/r/GA-LLM-D2408}{https://anonymous.4open.science/r/GA-LLM-D2408}.
\end{abstract}


\section{Introduction}


Next Point-of-Interest (nPOI) recommendation, which aims to predict a user’s next destination based on their historical movement data, plays a vital role in location-based services, enabling personalized applications such as tourism planning, restaurant discovery, and navigation. Traditional nPOI recommendation methods, while effective in modeling user preferences through collaborative filtering~\cite{FPMC} or graph-based approaches~\cite{STHGCN,GETNext}, often struggle to fully capture the semantic and contextual richness required for more complex and dynamic scenarios.


\begin{figure}[t]
    \centering
    \includegraphics[width=\linewidth]{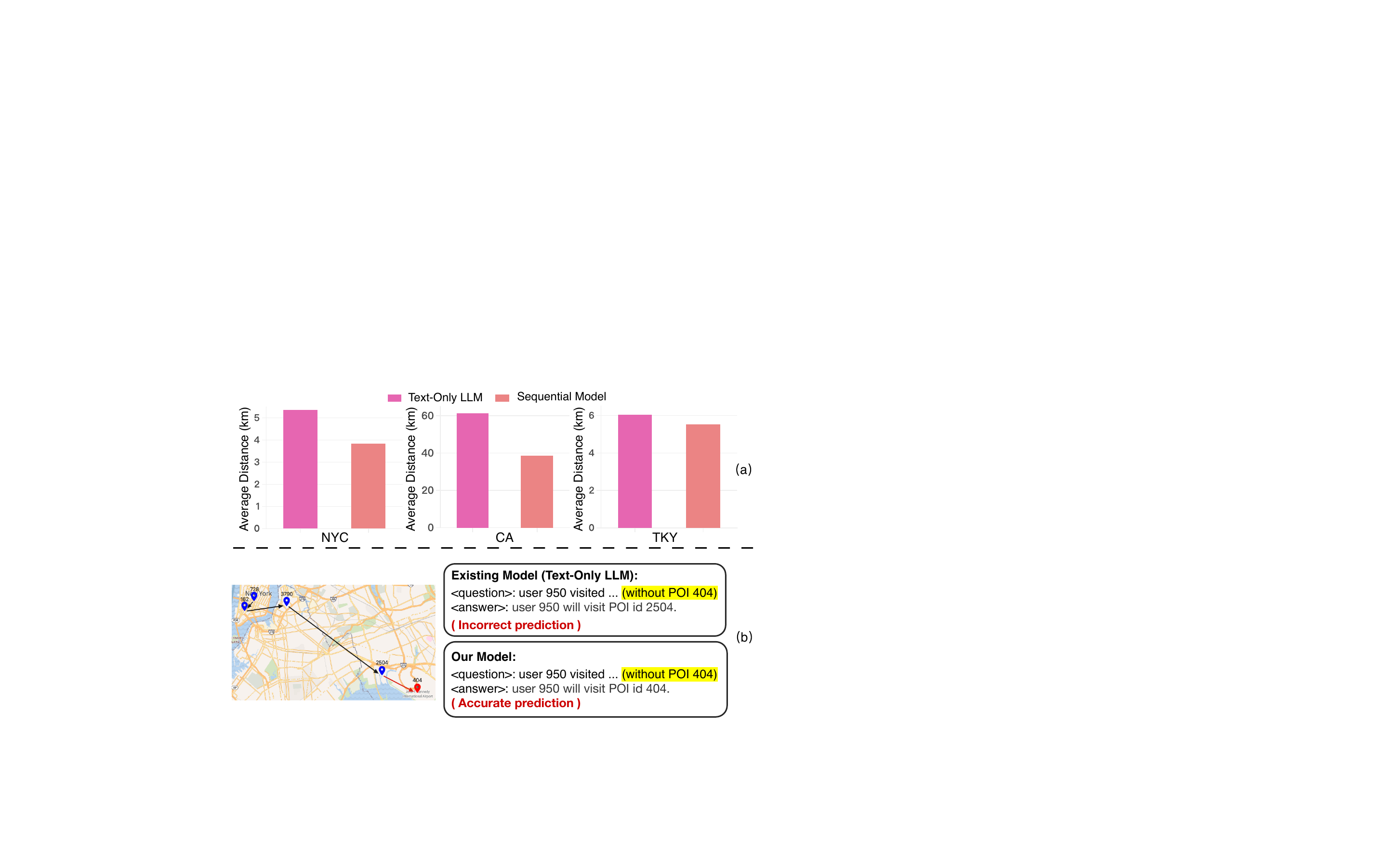}
    \caption{Illustration of text-only LLM limitations in POI recommendation. (a) Average error distances of incorrectly predicted POIs to their correct POIs for text-only LLMs versus sequential models on NYC, CA, and TKY datasets. (b) Case study where text-only LLM fails to predict POI 404 without input context, while our model leverages transition relations for accurate prediction.}
    \label{fig:case_study}
\end{figure}

The rise of Large Language Models (LLMs), with their unparalleled ability to model sequential dependencies and understand contextual semantics, has opened up exciting new opportunities for significantly advancing nPOI recommendation tasks. By leveraging LLMs~\cite{LLM4POI}, it becomes possible to bridge the gap between semantic modeling and diverse location-based data.
Recent works~\cite{GraphGPT,LLaRA,E4SRec} have shown the effectiveness of LLMs in recommendation tasks through their semantic understanding. However, applying LLMs to nPOI recommendation faces two key challenges.

\textbf{C1}: Limited spatial awareness. Existing methods~\cite{LLM4POI,E4SRec} often fail to capture spatial relationships, as LLMs are not inherently tailored to perceive geographic data like GPS coordinates~\cite{Pasquale2024City,Jonathan2023Gpt}. LLMs-based models may generate hallucinated POIs far from the ground truth due to a lack of consideration for spatial proximity. As shown in Figure~\ref{fig:case_study}(a), text-only LLMs produce higher error distances in incorrect predictions compared to sequential models (e.g., ROTAN) on three datasets, underscoring their limitations in effectively capturing spatial dependencies. When humans visit lots of POIs with unique GPS coordinates, these coordinates are critical for distinguishing their complex affinities. However, minor variations and low occurrence frequencies of GPS coordinates hinder LLMs' ability to model spatial relationships effectively. LLMs typically tokenize high-precision GPS coordinates, such as (40.8130989074707, -74.00180053710938), which have more than 10 decimal places. This process generates a large number of tokens (36 in this case), significantly escalating the computational cost and complicating the spatial relationship modeling. 

\textbf{C2}: Absence of POI transition relations. Traditional LLM-based methods rely mainly on textual information. Beyond the semantic relation, users' complete check-in trajectories hold rich potential POI transition knowledge. However, due to token limits in prompts, LLMs can only include couples of similar historical trajectories~\cite{LLM4POI}, making it challenging to model sequential dynamics, especially when the ground truth POI is absent from the historical trajectories. As illustrated in Figure~\ref{fig:case_study}(b), when POI 404 (Kennedy Airport) is not explicitly mentioned in the input, a baseline LLM model predicts a previously mentioned POI 2504 (AirTrain JFK station) as the next destination, failing to infer POI 404 through the POI-POI transition pattern. 



To address them, we propose the Geography-Aware Large Language Model (GA-LLM), integrating the Geographic Coordinate Injection Module (GCIM) and the Point-of-Interest Alignment Module (PAM). GCIM is designed to enhance the geographic awareness and adaptability of LLMs, transforming precise GPS coordinates into compact, efficient representations via hierarchical quadkey-based grid system, self-attention mechanisms and learnable Fourier transformations, respectively. PAM leverages global POI transition relationships, which can be extracted from sequential models \cite{CFPRec,GeoSAN} or graph-based models \cite{ROTAN,MTNet,GraphFlashback}, to align transition knowledge with the LLM’s semantic space. 






In this work, we make the following contributions:

\begin{itemize}
    \item We propose GA-LLM, a framework for next POI recommendation to mitigate hallucination issues of LLMs in spatial tasks. Especially, geographic coordinate injection
module is designed to enhance the geographic awareness of LLMs.  
    
    \item POI alignment module is proposed to incorporate POI transition relations into the semantic space of LLMs, enabling the model to capture sequential user movement patterns and address the limitations of text-only LLMs.
    
    \item Extensive experiments conducted on three public real-world datasets demonstrate that GA-LLM  outperforms existing  deep learning-based approaches and LLMs-based models in next POI recommendation tasks. Notably, it excels in addressing cold-start challenges and improving inference efficiency.
\end{itemize}

\begin{figure*}[t]
    \centering
    \includegraphics[width=\textwidth]{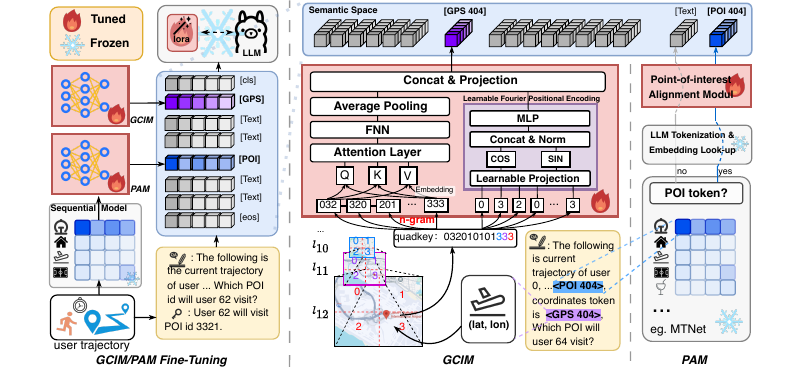}
    \caption{Overview of the GA-LLM framework. The left illustrates the workflow, where user trajectories are transformed into queries for LLM fine-tuning. The center highlights the GCIM module, encoding GPS data via a quadkey system and aligning it with the LLM’s semantic space. The right shows the PAM module, aligning POI embeddings from sequential models (e.g., MTNet) to the LLM’s semantic space.}
    \label{fig:framework}
\end{figure*}
\section{Related Work}

\subsection{Next POI Recommendation}



Early work on next POI recommendation \cite{DBLP:conf/ijcai/ChengYLK13,DBLP:conf/aaai/HeLLSC16} treats it as a sequential task using Markov chains and spatial constraints. With the rise of deep learning, RNN-based models gained popularity. HST-LSTM \cite{HST-LSTM} incorporates spatial-temporal factors into LSTM gates, while LSTPM \cite{LSTPM} combines geo-dilated LSTM and non-local operations to model short-term preferences. STAN \cite{STAN} uses multimodal embeddings with bi-layer attention, and CFPRec \cite{CFPRec} aligns multi-step planning with user preferences via attention. Graph-based methods advanced POI recommendation by modeling complex spatial-temporal relationships. GETNext \cite{GETNext} enhances POI representations with trajectory flow maps, while STHGCN \cite{STHGCN} and SNPM \cite{SNPM} leverage hypergraphs and dynamic neighbor graphs. ROTAN \cite{ROTAN} uses rotation-based temporal attention, and MTNet \cite{MTNet} models user preferences via mobility trees. However, these methods still face challenges in cold-start scenarios, particularly when there is limited user data, highlighting the need for more advanced models, such as LLMs, that are better suited to address these issues.

\subsection{LLMs for Recommender Systems}


Building on the success of LLMs in various domains \cite{CLIP,wav2vec}, recent research \cite{CoRAL,CoLLM,DBLP:journals/corr/abs-2405-20646} has explored their potential in recommender systems. Notable examples include GraphGPT \cite{GraphGPT}, aligning LLMs with graph-structured data via instruction tuning, and LLMRec \cite{LLMRec}, augmenting interaction graphs with LLM-generated side information. For sequential recommendation, SeCor \cite{SeCor} integrates semantic and collaborative embeddings, LLaRA \cite{LLaRA} employs hybrid prompting to align sequential user behavior, and ReLLa \cite{ReLLa} incorporates retrieval-based methods for long-sequence modeling. Semantic Convergence Framework \cite{AAAI2025} maps sparse collaborative semantics like ItemIDs into LLMs’ dense token space to improve scalability. For next POI recommendation, LLM4POI \cite{LLM4POI} uses trajectory prompting and key-query similarity to integrate heterogeneous LBSN data while addressing cold-start challenges. However, LLMs struggle with high-precision coordinates and fine-grained spatial relationships, limiting their application in location-based services. Our method incorporates spatial modeling and POI transitions to better handle geographic data and sequential dependencies.

\section{Problem Definition}

Next POI recommendation aims to predict the next location a user to visit based on their historical check-in records. Each check-in record is represented as a tuple \( q = (u, p, c, t, g) \), where \( u \) represents a user, \( p \) denotes the POI, \( c \) is the POI category (e.g., restaurant, park), \( t \) is the timestamp, and \( g = (\alpha, \beta) \) specifies the geographic coordinates of the POI. Given the historical trajectory of a user \( u \), \( T_u(t) = \{(p_1, c_1, t_1, g_1), \ldots, (p_k, c_k, t_k, g_k)\} \) up to time \( t \), the goal is to predict the next POI \( p_{k+1} \) that the user to visit at the subsequent timestamp \( t_{k+1} \). 
\section{Methods}


An idea for next POI recommendation is to leverage temporal and spatial trajectory information to model the likelihood of visiting candidate POIs. Accordingly, we propose the GA-LLM framework for this task.

\subsection{The GA-LLM Framework}

The GA-LLM framework, depicted in Figure~\ref{fig:framework}, integrates spatial and semantic modeling to address challenges in next POI recommendation. It encodes user trajectory data into structured prompts for LLM fine-tuning. The framework features two key modules: the Geographic Coordinate Injection Module (GCIM), which transforms high-precision GPS data into spatial representations, and the Point-of-Interest Alignment Module (PAM), which aligns graph-based POI embeddings with the LLM’s semantic space. Together, these modules enable the model to effectively capture spatial dependencies, global relationships, and sequential patterns.



\subsection{Geographic Coordinate Injection Module}
\label{sec:GCIM}

GCIM processes latitude and longitude coordinates. While LLMmove~\cite{LLMmove} considers distance for POI recommendation, our approach incorporates finer spatial dependencies. As shown in the middle of Figure~\ref{fig:framework}, GCIM maps GPS coordinates to hierarchical tiles using a unique Quadtree key (quadkey). It combines Fourier positional encoding with self-attention mechanisms to enrich quadkey representations, which are integrated into the model’s input, allowing LLMs to capture spatial dependencies more effectively.


\subsubsection{Semantic Prompt Construction}

We design structured prompts to encode POI details, including unique identifiers, categories, and geographic coordinates in a natural language format. For example:

\begin{tcolorbox}[colframe=black, colback=white, boxrule=0.5pt, arc=0pt, sharp corners]
\textit{“At \([t]\), user \([u]\) visited \(p\), represented as \(\textless \text{POI} \, [p] \textgreater\), with category \([c]\), located at \(\textless \text{GPS} \, [g] \textgreater\).”}
\end{tcolorbox}

GCIM processes \(\textless \text{GPS} \, [g] \textgreater\) to produce spatial representations, while PAM projects \(\textless \text{POI} \, [p] \textgreater\) into the LLM's semantic space. This structured design effectively integrates spatial dependencies and POI-specific information, enhancing downstream task performance.

\subsubsection{Quadkey-Based Geographic Encoding}

Inspired by the GeoEncoder proposed in GeoSAN~\cite{GeoSAN}, we adopt a quadtree-based grid encoding scheme to transform latitude and longitude into compact spatial representations. Quadkeys uniquely identify grid tiles at varying resolution levels. At a specified zoom level \( l \), the geographic space is projected into a 2D plane via the Mercator projection, mapping spherical coordinates \((\alpha, \beta)\) into Cartesian coordinates \((x, y)\) as follows:

\begin{equation}
x = \frac{\beta + 180}{360} \times 256 \times 2^l,
\end{equation}

\begin{equation}
y = \left( \frac{1}{2} - \frac{1}{4\pi} \log \frac{1 + \sin\left(\alpha \cdot \pi / 180\right)}{1 - \sin\left(\alpha \cdot \pi / 180\right)} \right) \times 256 \times 2^l,
\end{equation}
where, \( x \) and \( y \) are the positions of a GPS point on the grid at level \( l \). Each tile is assigned a quadkey by encoding the grid divisions into a base-4 string, where the string length corresponds to the zoom level \( l \). For a grid position, the resulting quadkey-derived representation \( S \) is defined as \( S = (s_1, s_2, \dots, s_L) \), where \( L \) represents the number of hierarchical levels in the quadtree. Each \( s_i \) \((i = 1, \dots, L)\) denotes the digit at the \( i \)-th level of the quadkey encoding, capturing the spatial structure at progressively finer resolutions. This transformation discretizes continuous GPS coordinates into grid cells, ensuring nearby locations share similar quadkeys and reducing sparsity, while \( S \) encapsulates the hierarchical spatial information of the quadkey.



To enrich the representations, overlapping \( n \)-grams (e.g., \( \{032, 320, 201, \dots\} \)) are generated from \( S \) to capture sequential patterns within the quadkey. Position encoding~\cite{DBLP:conf/icdm/KangM18} is then added to the \( n \)-gram embeddings, producing position-enhanced representations. A self-attention mechanism is applied to emphasize critical spatial patterns across hierarchical quadkey sequences and is formally defined as:
\[
\text{Attention}(Q, K, V) = \text{softmax}\left(\frac{QK^\top}{\sqrt{d_k}}\right)V,
\]
where \( Q, K, V \) are derived from the position-enhanced representations, and \( d_k \) is the dimensionality of the Key matrix. By dynamically assigning weights to \( n \)-grams, the self-attention mechanism captures both local spatial dependencies (from finer quadkeys) and global relationships (from coarser quadkeys), enhancing the model’s spatial representation.

\subsubsection{Learnable Fourier Positional Encoding}

This Fourier-based encoding introduces periodic transformations to the model through the representation of \( S \), enabling it to effectively capture fine-grained spatial patterns across multiple frequencies. By learning the projection matrix \( \mathbf{W_s} \), the encoding adapts to the spatial structure of the quadkey representation, thereby refining embeddings for downstream tasks. Fourier encoding preserves high-frequency variations to capture subtle local changes, while low-frequency patterns reflect broader geographic trends, ensuring comprehensive spatial relationship representation. 

Inspired by the learnable Fourier positional encoding approach proposed in \cite{LiSLHB21}, and building upon the methodology in \cite{M3PA}, which applies Fourier embeddings directly to latitude and longitude, we adapt this technique to operate on quadkey-derived representations. Quadkeys better capture hierarchical spatial structures by discretizing geographic space into grid tiles. The Fourier-based geographic embedding \( \mathbf{F}(S) \) is computed as follows:

\begin{equation}
\mathbf{F}(S) = \frac{1}{\sqrt{M}} \left[ \cos(S \cdot \mathbf{W_s}^T) \, || \, \sin(S \cdot \mathbf{W_s}^T) \right],
\label{eq:learnable_fourier}
\end{equation}
where \( S \in \mathbb{R}^L \) is the quadkey-derived grid position, \( M \) is the output embedding size, \( \mathbf{W_s} \in \mathbb{R}^{\frac{M}{2} \times L} \) is a learnable projection matrix initialized with a Gaussian distribution \( \mathbf{W_s} \sim N(0, \gamma^{-2}) \), and \( \gamma \) controls the spatial kernel width. The terms \( \cos(\cdot) \) and \( \sin(\cdot) \) denote element-wise cosine and sine functions, while \( || \) represents concatenation.

The resulting Fourier-based geographic embeddings \( \mathbf{F}(S) \) are concatenated with the \( n \)-gram/self-attention representations discussed earlier. These enriched representations serve complementary purposes: \( n \)-gram/self-attention captures hierarchical and sequential dependencies within the quadkey structure, while Fourier encoding captures global spatial patterns through detailed frequency decomposition. This seamless integration effectively combines local and global spatial dependencies, equipping the model to process complex geographic relationships and optimize overall performance.

\subsection{Point-of-Interest Alignment Module}

While GCIM enhances the model’s spatial understanding by capturing geographic information, it does not fully address challenges posed by Point-of-Interest (POI) interactions, such as user transition preferences~\cite{SNPM}. These interactions are often modeled using graph-based methods~\cite{STHGCN}, which produce low-dimensional embeddings that are not directly compatible with the high-dimensional semantic space of LLMs. To address this, we introduce the Point-of-Interest Alignment Module (PAM), which aligns graph-based POI embeddings with the LLM’s semantic space. Unlike methods like E4SRec~\cite{E4SRec}, which treat POIs as unique tokens and learn embeddings during fine-tuning, PAM bridges the gap between low-dimensional POI embeddings and the rich, contextual representations of LLMs, enabling the model to capture global relationships and user-specific behavioral patterns more effectively.

Specifically, PAM transforms POI embeddings generated by advanced graph-based models, such as MTNet \cite{MTNet}, STHGCN \cite{STHGCN}, and ROTAN \cite{ROTAN}, into compatible high-dimensional representations. Formally, let \(e_i \in \mathbb{R}^d\) denote the low-dimensional embedding of the \(i\)-th POI. PAM maps \(e_i\) into the LLM’s semantic space through a multi-layer perceptron (MLP), thereby producing \(h_i \in \mathbb{R}^D\). The alignment is formally defined as:
\begin{equation}
    h = h_\text{text} \cup h_i , \quad h_i = \text{PAM}(e_i),
\end{equation}
where \(h_\text{text}\) is the textual embedding derived from the LLM, and \(h \in \mathbb{R}^D\) combines textual and POI-specific information through concatenation. The MLP ensures that the POI embedding \(e_i\) is projected into a space that aligns with the LLM’s semantic features. Notably, GCIM (Section~\ref{sec:GCIM}) follows a similar process, where spatial representations derived from quadkey-based grid positions are also projected into the LLM's semantic space. This alignment ensures consistency between the POI-specific and spatial features, enabling a unified representation for downstream tasks.

By integrating spatial and transition-aware information into POI tokens, PAM enables the model to capture complex POI interactions, including user trajectories and movement preferences. Furthermore, when combined with the enriched spatial representations from GCIM, the unified feature representation allows the model to achieve a more comprehensive understanding of user behavior and spatial patterns, ultimately improving downstream task performance.

\section{Experiments}

In this section, we conduct extensive experiments on real-world datasets to evaluate the performance of the proposed GA-LLM by addressing the following research questions:

\begin{itemize}
    \item \textbf{RQ1:} Dose GA-LLM outperforms baseline methods in next POI recommendation?
    \item \textbf{RQ2:} What is the impact of key components (GCIM and PAM) on the performance of GA-LLM?
    \item \textbf{RQ3:} How do geographic coordinates, mitigation of LLM hallucinations via spatial encoding, and the inclusion of GCIM in cross-city cold-start scenarios impact the performance of next POI recommendation models?
    \item \textbf{RQ4:} How does treating POIs as unique tokens versus a mapping layer affect transition relations, and how does PAM improve modeling POI transitions?
    \item \textbf{RQ5:} How does GA-LLM achieve computational efficiency during the training and inference stages?
\end{itemize}

\subsection{Experimental Settings}

\noindent \textbf{Datasets and Metrics.} We evaluate our approach on three datasets: Foursquare-NYC, Foursquare-TKY~\cite{Foursquare}, and Gowalla-CA~\cite{Gowalla}, summarized in Table~\ref{tab:dataset_statistics}. Foursquare-NYC and Foursquare-TKY focus on New York City and Tokyo, while Gowalla-CA spans regions in California and Nevada. Each dataset provides user check-ins annotated with POI ID, category, location, and timestamp. Following~\cite{STHGCN}, we remove POIs and users with fewer than 10 check-ins and split the data chronologically into 80\% for training, 10\% for validation, and 10\% for testing. For evaluation, we use Top-1 Accuracy (\(Acc@1\)), which calculates the proportion of cases where the top-ranked POI matches the ground truth, framing recommendation as a question-answering task under the "extreme position bias" setting~\cite{LLM4POI,extreme_position_bias}.

\begin{table}[h]
    \centering
    \begin{tabular}{lcccc}
        \toprule
        \textbf{Dataset} & \textbf{\#user} & \textbf{\#poi} & \textbf{\#category} & \textbf{\#check-in}\\
        \midrule
        NYC        & 1,048           & 4,981          & 318                 & 103,941                          \\
        TKY        & 2,282           & 7,833          & 290                 & 405,000                   \\
        CA            & 3,957           & 9,690          & 296                 & 238,369                       \\
        \bottomrule
    \end{tabular}
    \caption{Statistics of the three experimental datasets.}
    \label{tab:dataset_statistics}
\end{table}


\noindent \textbf{Baselines.} Our experiments compare the proposed model with various baselines, including traditional methods like FPMC~\cite{FPMC} and PRME~\cite{PRME}, recurrent models such as LSTM~\cite{LSTM}, and advanced approaches incorporating spatiotemporal correlations, including STAN~\cite{STAN} and PLSPL~\cite{PLSPL}. Furthermore, we evaluate against graph-based methods like STGCN~\cite{STGCN} and hypergraph-based approaches such as STHGCN~\cite{STHGCN}. Transformer-based methods like GETNext~\cite{GETNext} and models addressing temporal dynamics, including ROTAN~\cite{ROTAN} and MTNet~\cite{MTNet}, are also considered. While many LLM-based approaches use a yes/no question-answer setup, which does not align with our task, we choose LLM4POI~\cite{LLM4POI} as the baseline because it directly predicts specific POI IDs, thus matching our setup more effectively.

\noindent \textbf{Implementation Details.} We use Llama-2-7b-longlora-32k~\cite{LongLoRA,Llama2} as our base model. Training uses a constant learning rate of \(2 \times 10^{-5}\), with 20 warm-up steps, no weight decay, and a batch size of 1 per GPU. The maximum sequence length is set to 32,768 tokens, and each dataset is fine-tuned for 3 epochs. All experiments run on servers equipped with Nvidia A800 GPUs.

\subsection{Main Results (RQ1)}

\begin{table}[t]
    \centering
    \begin{tabular}{lccc}
        \toprule
        Model       & \makecell{NYC \\ Acc@1} & \makecell{TKY \\ Acc@1} & \makecell{CA \\ Acc@1} \\
        \midrule
        FPMC        & 0.1003    & 0.0814    & 0.0383   \\
        LSTM        & 0.1305    & 0.1335    & 0.0665   \\
        PRME        & 0.1159    & 0.1052    & 0.0521   \\
        STGCN       & 0.1799    & 0.1716    & 0.0961   \\
        PLSPL       & 0.1917    & 0.1889    & 0.1072   \\
        STAN        & 0.2231    & 0.1963    & 0.1104   \\
        GETNext     & 0.2435    & 0.2254    & 0.1357   \\
        MTNext      & 0.2620    & 0.2575    & 0.1453   \\
        STHGCN      & 0.2734    & 0.2950    & 0.1730   \\
        ROTAN       & 0.3106    & 0.2458    & \underline{0.2199}   \\
        LLM4POI     & \underline{0.3372}    & \underline{0.3035}    & 0.2065    \\
        \midrule
        GA-LLM      & \textbf{0.3919} & \textbf{0.3482} & \textbf{0.2566} \\
        \textit{Improv.} & 16.19\% & 14.73\% & 16.69\% \\
        \bottomrule
    \end{tabular}
    \caption{Model performance. The overall best results are boldfaced, while the second-best results are underlined. Statistical significance testing shows that GA-LLM significantly outperforms the second-best baseline on all datasets, with $p$-value $\leq 0.05$.}
    \label{tab:performance_comparison}
\end{table}

We compare the performance of GA-LLM with various baselines on three datasets, as shown in Table~\ref{tab:performance_comparison}. GA-LLM achieves significant improvements in \(Acc@1\), surpassing the best baselines by 16.19\%, 14.73\%, and 16.69\% on the NYC, TKY, and CA datasets, respectively. These results highlight GA-LLM's ability to model diverse scenarios, from NYC's smaller and more structured dataset to CA's broader geographical coverage with higher data sparsity. Notably, while the improvement on the CA dataset over the best baseline ROTAN is 16.69\%, it is worth paying attention to LLM4POI, which also exploiting an LLM like GA-LLM dose. As Shown, GA-LLM achieves a much larger improvement of 24.26\% over LLM4POI. This demonstrates GA-LLM's effectiveness in overcoming the inherent limitations of text-only LLMs by incorporating geographic and POI-specific embeddings, which significantly enhance its ability to handle geographically extensive datasets with sparse spatial distributions. These findings underscore the strength of integrating spatial-temporal patterns and contextual relationships into the LLM framework, making GA-LLM particularly robust and adaptable for next POI recommendation tasks.

\subsection{Ablation Study (RQ2)}

To evaluate the impact of key components, we compare the performance of the full GA-LLM model, GA-LLM without GCIM, GA-LLM without PAM, and GA-LLM without the learnable Fourier transformation (w/o Fourier) across NYC, TKY, and CA datasets. As shown in Table~\ref{tab:ablation_study}, removing GCIM causes noticeable drops, particularly on datasets where spatial relationships are critical. Similarly, excluding PAM reduces accuracy across all datasets, highlighting its role in capturing POI transition relations. The ablation of the learnable Fourier transformation, a key component of GCIM, also results in performance degradation, demonstrating its importance in refining spatial representations.

\begin{figure}[t]
\centering
\includegraphics[width=1\linewidth]{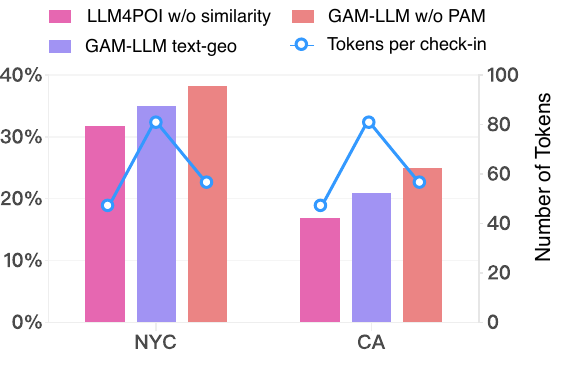}
\caption{Comparison of methods handling geographic coordinates.}
\label{fig:fig3_coor_analysis}
\end{figure}

\begin{table}[t]
    \centering
    \begin{tabular}{lccc}
        \toprule
        Model              & \makecell{NYC \\ Acc@1} & \makecell{TKY \\ Acc@1} & \makecell{CA \\ Acc@1} \\
        \midrule
        Full Model         & \textbf{0.3919} & \textbf{0.3482} & \textbf{0.2566} \\
        w/o GCIM           & 0.3729          & 0.3370          & 0.2402          \\
        w/o Fourier        & 0.3800          & 0.3435          & 0.2467          \\
        w/o PAM            & 0.3826          & 0.3429          & 0.2499          \\
        \bottomrule
    \end{tabular}
    \caption{Ablation study results on NYC, TKY, and CA datasets.}
    \label{tab:ablation_study}
\end{table}

\subsection{Geographic Information Analysis (RQ3)}

Geographic information is essential for next POI recommendation, addressing challenges in spatial modeling and cross-city cold-start scenarios. This section evaluates GA-LLM across three aspects: (1) geographic coordinates’ role in recommendation performance, (2) spatial encoding’s impact on reducing errors and LLM hallucinations, and (3) GCIM’s cross-city generalization capability.

\noindent \textbf{Geographic Coordinates in POI Recommendation.} We compare three methods of incorporating geographic information: (i) LLM4POI w/o similarity, excluding similar trajectories; (ii) GA-LLM text-geo, expressing geographic data as textual descriptions (e.g., “northwest 2.3 km from the previous POI”), increasing tokens; and (iii) GA-LLM w/o PAM, encoding latitude and longitude into compact representations via GCIM, avoiding verbose descriptions. As shown in Figure~\ref{fig:fig3_coor_analysis}, incorporating geographic information improves performance across all methods. GA-LLM w/o PAM achieves higher accuracy with fewer tokens, demonstrating GCIM’s efficiency and scalability for next POI recommendation.

\begin{figure}[t]
\centering
\includegraphics[width=1\linewidth]{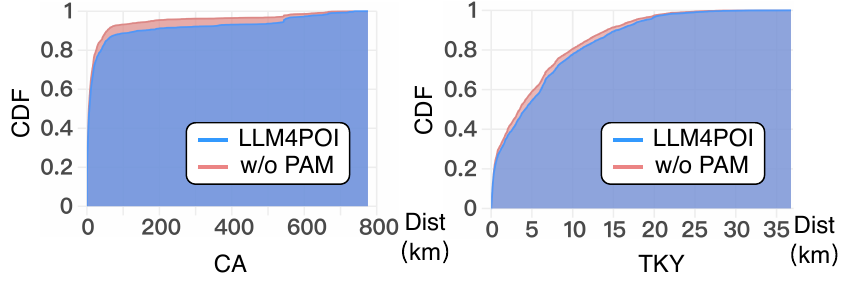}
\caption{Cumulative distribution function (CDF) of distances between incorrectly predicted and ground truth POIs on two datasets.}
\label{fig:fig4_distance_analysis}
\end{figure}

\begin{figure}[t]
\centering
\includegraphics[width=1\linewidth]{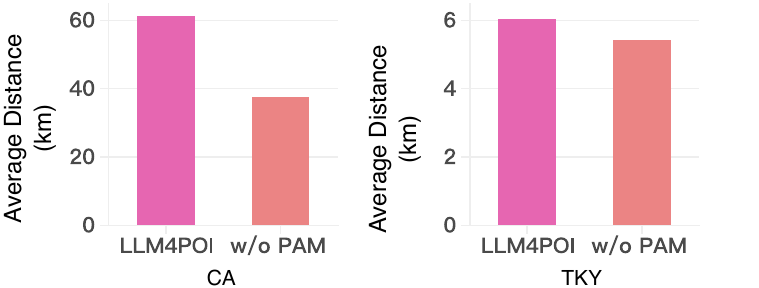}
\caption{Average error distances between incorrectly predicted and ground truth POIs on CA and TKY.}
\label{fig:fig5_average_distance_analysis}
\end{figure}

\begin{table}[b]
    \centering
    \begin{tabular}{@{}l@{\hskip 0pt}c@{\hskip 3pt}c c c@{}}
        \toprule
        Model & Trained on & \makecell{NYC \\ Acc@1} & \makecell{TKY \\ Acc@1} & \makecell{CA \\ Acc@1} \\
        \midrule
                         & NYC & 0.3372 & 0.2594 & 0.1885 \\
        LLM4POI & TKY & 0.3463 & 0.3035 & 0.1960 \\
                         & CA  & 0.3344 & 0.2600 & 0.2065 \\
        \midrule
                         & NYC & 0.3826 & 0.3018 & 0.2053 \\
        w/o PAM & TKY & 0.4059 & 0.3429 & 0.2273 \\
                         & CA  & 0.3670 & 0.3065 & 0.2499 \\
        \bottomrule
    \end{tabular}
    \caption{The models are LLM4POI and GA-LLM w/o PAM trained on one of the NYC, TKY, and CA datasets and evaluated on the rest.}
    \label{tab:cross_city_comparison}
\end{table}

\noindent \textbf{Mitigating LLM Hallucinations via Spatial Encoding.} Figures~\ref{fig:fig4_distance_analysis} and~\ref{fig:fig5_average_distance_analysis} demonstrate GCIM’s impact on reducing geographically implausible predictions. Figure~\ref{fig:fig4_distance_analysis} shows the CDF of error distances between predicted POIs and the ground truth, where GA-LLM w/o PAM results in smaller distances than LLM4POI. Figure~\ref{fig:fig5_average_distance_analysis} illustrates this with average distances reduced from 61.38 km to 37.63 km in the CA dataset. These results highlight GCIM’s ability to constrain prediction errors to plausible regions, effectively mitigating LLM hallucinations, especially in sparse datasets like CA.



\noindent \textbf{Cross-City Cold-Start Analysis.}  
Our GCIM module, a plug-in component of GA-LLM, enables effective integration of geographic information in cross-city cold-start scenarios. In this setting, a model trained on one city predicts POIs in another, a key challenge in location-based services. As shown in Table~\ref{tab:cross_city_comparison}, GA-LLM w/o PAM outperforms LLM4POI, achieving significant improvement in \(Acc@1\) for cross-city tasks. These results demonstrate GCIM’s ability to capture global geographic patterns and generalize across regions by emphasizing global features over local dependencies. Notably, performance for training on TKY and testing on NYC surpasses training and testing on NYC, due to TKY’s richer geographic diversity compared to NYC’s denser POI distribution.

\subsection{Sequential Transition Modeling (RQ4)}

This section analyzes the effectiveness of the Point-of-Interest Alignment Module (PAM) in capturing POI transition relations. The analysis is divided into two parts: (1) evaluating the strategy of treating POIs as new tokens and its limitations, and (2) demonstrating the benefits of PAM in improving accuracy and capturing transition relations.

\noindent \textbf{Evaluating POI Tokens and Transition Injection.}
We compare our approach to E4SRec~\cite{E4SRec}, which treats each POI as a unique token in the vocabulary, initializing POI embeddings with pre-trained representations from MTNet and fine-tuning during training. In contrast, our PAM-based method employs a projector layer during LLM fine-tuning to align POI representations with the LLM’s semantic space, avoiding reliance on individual POI-specific embeddings. As shown in Table~\ref{tab:token_comparison}, the token-based approach performs better in CA, likely due to traditional POI embeddings capturing transition relations in sparse datasets. However, it struggles in NYC, where the higher density and proximity of POIs make distinguishing embeddings during initialization challenging. This highlights a limitation of the tokenization strategy in dense urban environments. In contrast, our PAM-based approach overcomes these challenges by leveraging semantic mappings to model transition relations, delivering consistently strong performance across datasets.

\begin{figure}[t]
\centering
\includegraphics[width=1\linewidth]{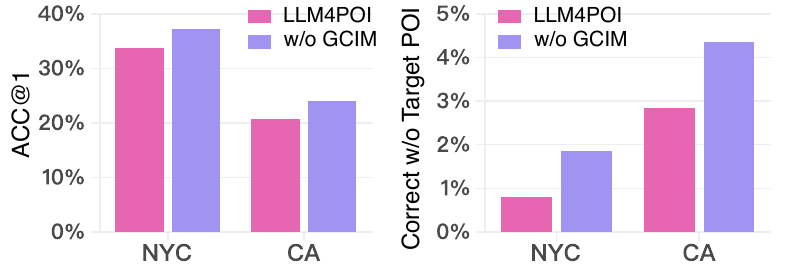}
\caption{Proportion of successful predictions where the target POI does not appear in the input question, for LLM4POI and GA-LLM w/o GCIM across NYC and CA datasets.}
\label{fig:fig6_pam_analysis}
\end{figure}

\begin{table}[t]
    \centering
    \begin{tabular}{lcc}
        \toprule
        Model               & \makecell{NYC \\ Acc@1} & \makecell{CA \\ Acc@1} \\
        \midrule
        LLM4POI             & 0.3372        & 0.2065       \\
        token-based       & 0.3389        & 0.2226       \\
        w/o GCIM     & \textbf{0.3729} & \textbf{0.2402} \\
        \bottomrule
    \end{tabular}
    \caption{Performance comparison between token-based POI representation and our PAM-based approach.}
    \label{tab:token_comparison}
\end{table}

\noindent \textbf{Effectiveness of Transition Preference Injection.} Capturing POI transition relations is critical for predicting target POIs that do not explicitly appear in the input question. In the test datasets, 75.8\% of target POIs in NYC, 73.5\% in TKY, and 55.6\% in CA appear in the input question. Text-only LLMs like LLM4POI rely heavily on the presence of the target POI in the input, limiting performance when this context is missing. Our PAM module addresses this limitation by capturing transition relations through contextualized embeddings. As shown in Figure~\ref{fig:fig6_pam_analysis}, PAM improves the correct predictions where the target POI is absent from the input, increasing ACC@1 from 2.85\% to 4.36\% in CA, demonstrating its effectiveness. Furthermore, PAM is a flexible and generalizable component compatible with various sequential models. While our experiments use MTNet, additional results on other models in the supplementary materials demonstrate its adaptability across scenarios.

\subsection{Model Efficiency Study (RQ5)}
The study evaluates the computational efficiency of GA-LLM during training and inference stages, compared to LLM4POI.

\begin{figure}[t]
\centering
\includegraphics[width=1\linewidth]{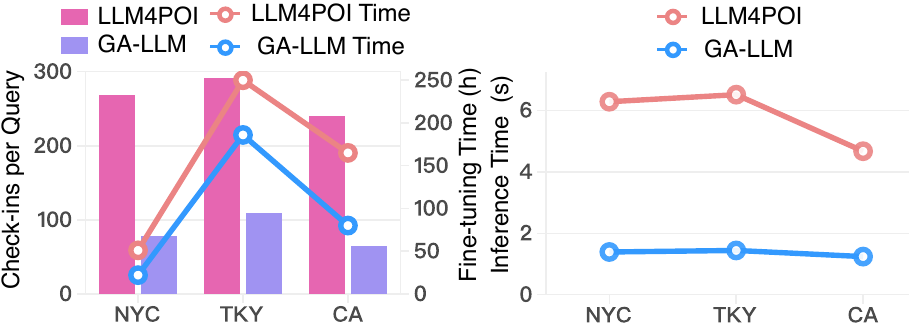}
\caption{Efficiency analysis of GA-LLM and LLM4POI. Left: Fine-tuning time and average check-ins per query. Right: Per-query inference time across datasets.}
\label{fig:fig7_efficiency_study}
\end{figure}

\noindent \textbf{Training Efficiency Analysis.}
The left side of Figure~\ref{fig:fig7_efficiency_study} compares the training efficiency of GA-LLM and LLM4POI, where LLM4POI uses 200 historical check-ins per query. GA-LLM achieves shorter fine-tuning times across datasets by processing fewer check-ins, reducing token consumption. For fairness, we focus solely on the final fine-tuning stage, excluding pre-training for GCIM and PAM.

\noindent \textbf{Inference Efficiency Analysis.}
The right side of Figure~\ref{fig:fig7_efficiency_study} shows that GA-LLM achieves faster inference time per query than LLM4POI, primarily due to processing fewer check-ins per query. The components in GA-LLM, i.e., GCIM and PAM, introduce minimal overhead, ensuring both efficient inference and consistently high accuracy.


\section{Conclusion}

In this paper, we propose GA-LLM, a framework that integrates geographic and sequential transition modeling into large language models for next POI recommendation. In GA-LLM, the Geographic Coordinate Injection Module (GCIM) converts precise GPS data into compact spatial representations, capturing geographic patterns across multiple granularities and perspectives. The Point-of-Interest Alignment Module (PAM) models POI transition relations, leveraging contextual embeddings to address text-only LLM limitations on exploring unvisited POIs. Extensive experiments confirm the effectiveness and efficiency of integrating geography-aware representations, POI transition knowledge and the semantic understanding capabilities of LLMs. While GA-LLM shows strong performance in accuracy and adaptability across datasets, challenges remain in sparse datasets and evolving user behaviors. Future work is planned to align user behavior modalities, such as visit times, into the LLMs' semantic space to further enhance geographic understanding and user preferences.

\clearpage 
\bibliographystyle{named}
\bibliography{References}

\begin{thebibliography}{}

\bibitem[\protect\citeauthoryear{Baevski \bgroup \em et al.\egroup }{2020}]{wav2vec}
Alexei Baevski, Yuhao Zhou, Abdelrahman Mohamed, and Michael Auli.
\newblock wav2vec 2.0: {A} framework for self-supervised learning of speech representations.
\newblock In {\em NeurIPS}, 2020.

\bibitem[\protect\citeauthoryear{Balsebre \bgroup \em et al.\egroup }{2024}]{Pasquale2024City}
Pasquale Balsebre, Weiming Huang, Gao Cong, and Yi~Li.
\newblock City foundation models for learning general purpose representations from openstreetmap.
\newblock In {\em CIKM}, pages 87--97. {ACM}, 2024.

\bibitem[\protect\citeauthoryear{Chen \bgroup \em et al.\egroup }{2024}]{LongLoRA}
Yukang Chen, Shengju Qian, Haotian Tang, Xin Lai, Zhijian Liu, Song Han, and Jiaya Jia.
\newblock Longlora: Efficient fine-tuning of long-context large language models.
\newblock In {\em {ICLR}}. OpenReview.net, 2024.

\bibitem[\protect\citeauthoryear{Cheng \bgroup \em et al.\egroup }{2013}]{DBLP:conf/ijcai/ChengYLK13}
Chen Cheng, Haiqin Yang, Michael~R. Lyu, and Irwin King.
\newblock Where you like to go next: Successive point-of-interest recommendation.
\newblock In {\em IJCAI}, pages 2605--2611. {IJCAI/AAAI}, 2013.

\bibitem[\protect\citeauthoryear{Cho \bgroup \em et al.\egroup }{2011}]{Gowalla}
Eunjoon Cho, Seth~A. Myers, and Jure Leskovec.
\newblock Friendship and mobility: user movement in location-based social networks.
\newblock In {\em Proceedings of the 17th {ACM} {SIGKDD} International Conference on Knowledge Discovery and Data Mining}, pages 1082--1090. {ACM}, 2011.

\bibitem[\protect\citeauthoryear{Fairstein \bgroup \em et al.\egroup }{2022}]{extreme_position_bias}
Yaron Fairstein, Elad Haramaty, Arnon Lazerson, and Liane Lewin{-}Eytan.
\newblock External evaluation of ranking models under extreme position-bias.
\newblock In {\em {WSDM} '22: The Fifteenth {ACM} International Conference on Web Search and Data Mining}, pages 252--261. {ACM}, 2022.

\bibitem[\protect\citeauthoryear{Feng \bgroup \em et al.\egroup }{2015}]{PRME}
Shanshan Feng, Xutao Li, Yifeng Zeng, Gao Cong, Yeow~Meng Chee, and Quan Yuan.
\newblock Personalized ranking metric embedding for next new {POI} recommendation.
\newblock In {\em {IJCAI}}, pages 2069--2075. {AAAI} Press, 2015.

\bibitem[\protect\citeauthoryear{Feng \bgroup \em et al.\egroup }{2024a}]{LLMmove}
Shanshan Feng, Haoming Lyu, Caishun Chen, and Yew-Soon Ong.
\newblock Where to move next: Zero-shot generalization of llms for next poi recommendation, 2024.

\bibitem[\protect\citeauthoryear{Feng \bgroup \em et al.\egroup }{2024b}]{ROTAN}
Shanshan Feng, Feiyu Meng, Lisi Chen, Shuo Shang, and Yew~Soon Ong.
\newblock {ROTAN:} {A} rotation-based temporal attention network for time-specific next {POI} recommendation.
\newblock In {\em Proceedings of the 30th {ACM} {SIGKDD} Conference on Knowledge Discovery and Data Mining}, pages 759--770. {ACM}, 2024.

\bibitem[\protect\citeauthoryear{He \bgroup \em et al.\egroup }{2016}]{DBLP:conf/aaai/HeLLSC16}
Jing He, Xin Li, Lejian Liao, Dandan Song, and William~K. Cheung.
\newblock Inferring a personalized next point-of-interest recommendation model with latent behavior patterns.
\newblock In {\em Proceedings of the Thirtieth {AAAI} Conference on Artificial Intelligence}, pages 137--143. {AAAI} Press, 2016.

\bibitem[\protect\citeauthoryear{Hochreiter and Schmidhuber}{1997}]{LSTM}
Sepp Hochreiter and J{\"{u}}rgen Schmidhuber.
\newblock Long short-term memory.
\newblock {\em Neural Comput.}, 9(8):1735--1780, 1997.

\bibitem[\protect\citeauthoryear{Huang \bgroup \em et al.\egroup }{2024}]{MTNet}
Tianhao Huang, Xuan Pan, Xiangrui Cai, Ying Zhang, and Xiaojie Yuan.
\newblock Learning time slot preferences via mobility tree for next {POI} recommendation.
\newblock In {\em Thirty-Eighth {AAAI} Conference on Artificial Intelligence}, pages 8535--8543. {AAAI} Press, 2024.

\bibitem[\protect\citeauthoryear{Kang}{2018}]{DBLP:conf/icdm/KangM18}
Wang{-}Cheng Kang.
\newblock Self-attentive sequential recommendation.
\newblock In {\em {ICDM}}, pages 197--206. {IEEE} Computer Society, 2018.

\bibitem[\protect\citeauthoryear{Kong and Wu}{2018}]{HST-LSTM}
Dejiang Kong and Fei Wu.
\newblock {HST-LSTM:} {A} hierarchical spatial-temporal long-short term memory network for location prediction.
\newblock In {\em IJCAI}, pages 2341--2347. ijcai.org, 2018.

\bibitem[\protect\citeauthoryear{Li \bgroup \em et al.\egroup }{2021}]{LiSLHB21}
Yang Li, Si~Si, Gang Li, Cho{-}Jui Hsieh, and Samy Bengio.
\newblock Learnable fourier features for multi-dimensional spatial positional encoding.
\newblock In {\em NeurIPS}, pages 15816--15829, 2021.

\bibitem[\protect\citeauthoryear{Li \bgroup \em et al.\egroup }{2023}]{E4SRec}
Xinhang Li, Chong Chen, Xiangyu Zhao, Yong Zhang, and Chunxiao Xing.
\newblock E4srec: An elegant effective efficient extensible solution of large language models for sequential recommendation.
\newblock {\em CoRR}, abs/2312.02443, 2023.

\bibitem[\protect\citeauthoryear{Li \bgroup \em et al.\egroup }{2024a}]{AAAI2025}
Guanghan Li, Xun Zhang, Yufei Zhang, Yifan Yin, Guojun Yin, and Wei Lin.
\newblock Semantic convergence: Harmonizing recommender systems via two-stage alignment and behavioral semantic tokenization, 2024.

\bibitem[\protect\citeauthoryear{Li \bgroup \em et al.\egroup }{2024b}]{LLM4POI}
Peibo Li, Maarten de~Rijke, Hao Xue, Shuang Ao, Yang Song, and Flora~D. Salim.
\newblock Large language models for next point-of-interest recommendation.
\newblock In {\em {SIGIR}}, pages 1463--1472. {ACM}, 2024.

\bibitem[\protect\citeauthoryear{Lian \bgroup \em et al.\egroup }{2020}]{GeoSAN}
Defu Lian, Yongji Wu, Yong Ge, Xing Xie, and Enhong Chen.
\newblock Geography-aware sequential location recommendation.
\newblock In {\em The 26th {ACM} {SIGKDD} Conference on Knowledge Discovery and Data Mining}, pages 2009--2019. {ACM}, 2020.

\bibitem[\protect\citeauthoryear{Liao \bgroup \em et al.\egroup }{2024}]{LLaRA}
Jiayi Liao, Sihang Li, Zhengyi Yang, Jiancan Wu, Yancheng Yuan, Xiang Wang, and Xiangnan He.
\newblock Llara: Large language-recommendation assistant.
\newblock In {\em SIGIR}, pages 1785--1795. {ACM}, 2024.

\bibitem[\protect\citeauthoryear{Lin \bgroup \em et al.\egroup }{2024}]{ReLLa}
Jianghao Lin, Rong Shan, Chenxu Zhu, Kounianhua Du, Bo~Chen, Shigang Quan, Ruiming Tang, Yong Yu, and Weinan Zhang.
\newblock Rella: Retrieval-enhanced large language models for lifelong sequential behavior comprehension in recommendation.
\newblock In {\em Proceedings of the {ACM} on Web Conference 2024}, pages 3497--3508. {ACM}, 2024.

\bibitem[\protect\citeauthoryear{Liu \bgroup \em et al.\egroup }{2024}]{DBLP:journals/corr/abs-2405-20646}
Qidong Liu, Xian Wu, Xiangyu Zhao, Yejing Wang, Zijian Zhang, Feng Tian, and Yefeng Zheng.
\newblock Large language models enhanced sequential recommendation for long-tail user and item.
\newblock {\em CoRR}, abs/2405.20646, 2024.

\bibitem[\protect\citeauthoryear{Luo \bgroup \em et al.\egroup }{2021}]{STAN}
Yingtao Luo, Qiang Liu, and Zhaocheng Liu.
\newblock {STAN:} spatio-temporal attention network for next location recommendation.
\newblock In {\em WWW}, pages 2177--2185. {ACM} / {IW3C2}, 2021.

\bibitem[\protect\citeauthoryear{Radford \bgroup \em et al.\egroup }{2021}]{CLIP}
Alec Radford, Jong~Wook Kim, Chris Hallacy, Aditya Ramesh, Gabriel Goh, Sandhini Agarwal, Girish Sastry, Amanda Askell, Pamela Mishkin, Jack Clark, Gretchen Krueger, and Ilya Sutskever.
\newblock Learning transferable visual models from natural language supervision.
\newblock In {\em Proceedings of the 38th International Conference on Machine Learning}, volume 139 of {\em Proceedings of Machine Learning Research}, pages 8748--8763. {PMLR}, 2021.

\bibitem[\protect\citeauthoryear{Rao \bgroup \em et al.\egroup }{2022}]{GraphFlashback}
Xuan Rao, Lisi Chen, Yong Liu, Shuo Shang, Bin Yao, and Peng Han.
\newblock Graph-flashback network for next location recommendation.
\newblock In {\em The 28th {ACM} {SIGKDD} Conference on Knowledge Discovery and Data Mining}, pages 1463--1471. {ACM}, 2022.

\bibitem[\protect\citeauthoryear{Rendle \bgroup \em et al.\egroup }{2010}]{FPMC}
Steffen Rendle, Christoph Freudenthaler, and Lars Schmidt{-}Thieme.
\newblock Factorizing personalized markov chains for next-basket recommendation.
\newblock In {\em Proceedings of the 19th International Conference on World Wide Web}, pages 811--820. {ACM}, 2010.

\bibitem[\protect\citeauthoryear{Roberts \bgroup \em et al.\egroup }{2023}]{Jonathan2023Gpt}
Jonathan Roberts, Timo L{\"{u}}ddecke, Sowmen Das, Kai Han, and Samuel Albanie.
\newblock {GPT4GEO:} how a language model sees the world's geography.
\newblock {\em CoRR}, abs/2306.00020, 2023.

\bibitem[\protect\citeauthoryear{Sun \bgroup \em et al.\egroup }{2020}]{LSTPM}
Ke~Sun, Tieyun Qian, Tong Chen, Yile Liang, Quoc Viet~Hung Nguyen, and Hongzhi Yin.
\newblock Where to go next: Modeling long- and short-term user preferences for point-of-interest recommendation.
\newblock In {\em AAAI}, pages 214--221. {AAAI} Press, 2020.

\bibitem[\protect\citeauthoryear{Tang \bgroup \em et al.\egroup }{2024}]{GraphGPT}
Jiabin Tang, Yuhao Yang, Wei Wei, Lei Shi, Lixin Su, Suqi Cheng, Dawei Yin, and Chao Huang.
\newblock Graphgpt: Graph instruction tuning for large language models.
\newblock In {\em SIGIR}, pages 491--500. {ACM}, 2024.

\bibitem[\protect\citeauthoryear{Touvron \bgroup \em et al.\egroup }{2023}]{Llama2}
Hugo Touvron, Louis Martin, Kevin Stone, Peter Albert, Amjad Almahairi, and et.
\newblock Llama 2: Open foundation and fine-tuned chat models.
\newblock {\em CoRR}, abs/2307.09288, 2023.

\bibitem[\protect\citeauthoryear{Wang \bgroup \em et al.\egroup }{2024}]{SeCor}
Shirui Wang, Bohan Xie, Ling Ding, Xiaoying Gao, Jianting Chen, and Yang Xiang.
\newblock Secor: Aligning semantic and collaborative representations by large language models for next-point-of-interest recommendations.
\newblock In {\em Proceedings of the 18th {ACM} Conference on Recommender Systems, RecSys}, pages 1--11. {ACM}, 2024.

\bibitem[\protect\citeauthoryear{Wei \bgroup \em et al.\egroup }{2024}]{LLMRec}
Wei Wei, Xubin Ren, Jiabin Tang, Qinyong Wang, Lixin Su, Suqi Cheng, Junfeng Wang, Dawei Yin, and Chao Huang.
\newblock Llmrec: Large language models with graph augmentation for recommendation.
\newblock In {\em Proceedings of the 17th {ACM} International Conference on Web Search and Data Mining}, pages 806--815. {ACM}, 2024.

\bibitem[\protect\citeauthoryear{Wu \bgroup \em et al.\egroup }{2022}]{PLSPL}
Yuxia Wu, Ke~Li, Guoshuai Zhao, and Xueming Qian.
\newblock Personalized long- and short-term preference learning for next {POI} recommendation.
\newblock {\em {IEEE} Trans. Knowl. Data Eng.}, 34(4):1944--1957, 2022.

\bibitem[\protect\citeauthoryear{Wu \bgroup \em et al.\egroup }{2024}]{CoRAL}
Junda Wu, Cheng{-}Chun Chang, Tong Yu, Zhankui He, Jianing Wang, Yupeng Hou, and Julian~J. McAuley.
\newblock Coral: Collaborative retrieval-augmented large language models improve long-tail recommendation.
\newblock In {\em Proceedings of the 30th {ACM} {SIGKDD} Conference on Knowledge Discovery and Data Mining}, pages 3391--3401. {ACM}, 2024.

\bibitem[\protect\citeauthoryear{Yan \bgroup \em et al.\egroup }{2023}]{STHGCN}
Xiaodong Yan, Tengwei Song, Yifeng Jiao, Jianshan He, Jiaotuan Wang, Ruopeng Li, and Wei Chu.
\newblock Spatio-temporal hypergraph learning for next {POI} recommendation.
\newblock In {\em SIGIR}, pages 403--412. {ACM}, 2023.

\bibitem[\protect\citeauthoryear{Yang \bgroup \em et al.\egroup }{2015}]{Foursquare}
Dingqi Yang, Daqing Zhang, Vincent~W. Zheng, and Zhiyong Yu.
\newblock Modeling user activity preference by leveraging user spatial temporal characteristics in lbsns.
\newblock {\em {IEEE} Trans. Syst. Man Cybern. Syst.}, 45(1):129--142, 2015.

\bibitem[\protect\citeauthoryear{Yang \bgroup \em et al.\egroup }{2022}]{GETNext}
Song Yang, Jiamou Liu, and Kaiqi Zhao.
\newblock Getnext: Trajectory flow map enhanced transformer for next {POI} recommendation.
\newblock In {\em SIGIR}, pages 1144--1153. {ACM}, 2022.

\bibitem[\protect\citeauthoryear{Yin \bgroup \em et al.\egroup }{2023}]{SNPM}
Feiyu Yin, Yong Liu, Zhiqi Shen, Lisi Chen, Shuo Shang, and Peng Han.
\newblock Next {POI} recommendation with dynamic graph and explicit dependency.
\newblock In {\em AAAI}, pages 4827--4834. {AAAI} Press, 2023.

\bibitem[\protect\citeauthoryear{Zhang \bgroup \em et al.\egroup }{2022}]{CFPRec}
Lu~Zhang, Zhu Sun, Ziqing Wu, Jie Zhang, Yew~Soon Ong, and Xinghua Qu.
\newblock Next point-of-interest recommendation with inferring multi-step future preferences.
\newblock In {\em IJCAI}, pages 3751--3757. ijcai.org, 2022.

\bibitem[\protect\citeauthoryear{Zhang \bgroup \em et al.\egroup }{2023}]{CoLLM}
Yang Zhang, Fuli Feng, Jizhi Zhang, Keqin Bao, Qifan Wang, and Xiangnan He.
\newblock Collm: Integrating collaborative embeddings into large language models for recommendation.
\newblock {\em CoRR}, abs/2310.19488, 2023.

\bibitem[\protect\citeauthoryear{Zhang \bgroup \em et al.\egroup }{2024}]{M3PA}
Dabin Zhang, Meng Chen, Weiming Huang, Yongshun Gong, and Kai Zhao.
\newblock Exploring urban semantics: {A} multimodal model for {POI} semantic annotation with street view images and place names.
\newblock In {\em IJCAI}, pages 2533--2541. ijcai.org, 2024.

\bibitem[\protect\citeauthoryear{Zhao \bgroup \em et al.\egroup }{2022}]{STGCN}
Pengpeng Zhao, Anjing Luo, Yanchi Liu, Jiajie Xu, Zhixu Li, Fuzhen Zhuang, Victor~S. Sheng, and Xiaofang Zhou.
\newblock Where to go next: {A} spatio-temporal gated network for next {POI} recommendation.
\newblock {\em {IEEE} Trans. Knowl. Data Eng.}, 34(5):2512--2524, 2022.

\end{thebibliography}

\end{document}





\appendix
\section*{APPENDIX}

\subsection*{A. Analysis of POI Embedding Sources in PAM}

The Point-of-Interest Alignment Module (PAM) is a flexible component that can align POI embeddings from various sequence models to the semantic space of large language models (LLMs). In our main experiments, we use POI embeddings of dimension 128 derived from the MTNet sequence model. To further evaluate the robustness of PAM, we compare its performance with POI embeddings from other foundational models, including STHGCN and ROTAN. Since ROTAN incorporates a rotation-based attention mechanism, we utilize the POI embeddings obtained from its first-stage pre-training, which also have a dimension of 128. 

We now present a comparison of the performance of GA-LLM using POI embeddings from these different sequence models, specifically focusing on the impact on accuracy on the NYC dataset. The following Table~\ref{tab:poembeddings} shows the performance comparison of GA-LLM with MTNet, STHGCN, and ROTAN embeddings.

\setcounter{table}{4} 
\begin{table}[h]
    \centering
    \begin{tabular}{lccc}
        \toprule
        \textbf{Model} & \makecell{NYC \\ Acc@1} \\
        \midrule
        GA-LLM-\textit{MTNet}     & 0.3919 \\
        GA-LLM-\textit{STHGCN}    & 0.3882 \\
        GA-LLM-\textit{ROTAN}     & 0.3853 \\
        \bottomrule
    \end{tabular}
    \caption{Comparison of performance using different POI embeddings in PAM on the NYC dataset.}
    \label{tab:poembeddings}
\end{table}

\subsection*{B. Alignment Layer Analysis in PAM}

In this section, we explore the impact of the number of alignment layers in the Point-of-Interest Alignment Module (PAM) on the model's performance. Specifically, we compare the performance of PAM with 1 layer versus 2 layers across the NYC and CA datasets. The results indicate that there is no significant performance difference between using 1 layer and 2 layers. However, the use of a single alignment layer offers a notable advantage in terms of reducing the number of parameters, making it a more efficient choice. Therefore, we adopt the 1-layer configuration for PAM in our main experiments.

The following table presents the performance comparison of PAM with 1 and 2 alignment layers on the NYC and CA datasets.

\begin{table}[h]
    \centering
    \begin{tabular}{lcc}
        \toprule
        \textbf{Model} & \makecell{NYC \\ Acc@1} & \makecell{CA \\ Acc@1} \\
        \midrule
        PAM (1 layer)  & 0.3729 & 0.2402 \\
        PAM (2 layers) & 0.3700 & 0.2399 \\
        \bottomrule
    \end{tabular}
    \caption{Performance comparison of PAM with 1 and 2 alignment layers on the NYC and CA datasets.}
    \label{tab:pam_alignment_layers}
\end{table}

\subsection*{C. Quadkey-Based Parameter Settings}

In the Quadkey-Based Geographic Encoding, the parameter \( L \) specifies the number of hierarchical levels in the quadtree, which directly determines the resolution of the spatial representation. In our experiments, we set \( L = 25 \), corresponding to a fine-grained grid resolution. This value ensures a balance between capturing detailed spatial structures and maintaining computational efficiency. A higher \( L \) results in smaller grid tiles, allowing for greater differentiation of nearby locations, while a lower \( L \) creates coarser tiles that may lose fine spatial details. The choice of \( L = 25 \) is motivated by the need to adequately represent dense urban areas, where high spatial granularity is critical for modeling user movements and POI distributions effectively.